\documentclass[prb,twocolumn,floatfix,preprintnumbers,amsmath,amssymb,
superscriptaddress, a4paper]{revtex4-2}
\usepackage{geometry}
\usepackage{float}
\usepackage{graphicx}% Include figure files
\usepackage{dcolumn}% Align table columns on decimal point
\usepackage{bm}% bold math
\usepackage{color}
\usepackage{xspace}
\usepackage{amsmath}
\usepackage{appendix}
\usepackage{multirow} 
\usepackage{subfigure}

\usepackage{textcomp} % Use with pdfLaTeX

\graphicspath{{./}{figure/}}
\usepackage[colorlinks,plainpages=false,linkcolor=blue,urlcolor=blue,citecolor=blue,pdfpagemode=UseNone,pdfstartview=FitBH]{hyperref}
\usepackage{dcolumn,graphicx,color,booktabs,microtype,afterpage}
\usepackage[charter,greekuppercase=italicized]{mathdesign}
\newcommand{\chemDIMPY}{(C$_7$H$_{10}$N$_2$)$_2$CuBr$_4$}
\newcommand{\chemCPA}{(C$_5$H$_9$NH$_3$)$_2$CuBr$_4$}
\newcommand{\chemCPAD}{(C$_5$D$_9$ND$_3$)$_2$CuBr$_4$}
\newcommand{\CPA}{Cu-CPA}
\begin{document}

\title{\chemCPA: a metal-organic two-ladder quantum magnet}

\author{J.~Philippe}
\email{jonas.philippe@psi.ch}
\affiliation{PSI Center for Neutron and Muon Sciences, Paul Scherrer Institut, CH-5232 Villigen-PSI, Switzerland}
\affiliation{Physik-Institut, Universit\"{a}t Z\"{u}rich, Winterthurerstrasse 190, CH-8057 Z\"{u}rich, Switzerland}

\author{F.~Elson}
\affiliation{Department of Applied Physics, KTH Royal Institute of Technology, SE-106 91 Stockholm, Sweden}

\author{N.~P.~M.~Casati}
\email{nicola.casati@psi.ch}
\affiliation{PSI Center for Photon Sciences, Paul Scherrer Institut, CH-5232 Villigen-PSI, Switzerland}

\author{S.~Sanz}
\affiliation{Peter Grünberg Institute, Electronic Properties (PGI-6), Forschungszentrum Jülich, 52425 Jülich, Germany}

\author{M.~Metzelaars}
\affiliation{Peter Grünberg Institute, Electronic Properties (PGI-6), Forschungszentrum Jülich, 52425 Jülich, Germany}
\affiliation{Institute of Inorganic Chemistry, RWTH Aachen University, 52056 Aachen, Germany}

\author{O.~Shliakhtun}
\affiliation{PSI Center for Neutron and Muon Sciences, Paul Scherrer Institut, CH-5232 Villigen-PSI, Switzerland}
\affiliation{Physik-Institut, Universit\"{a}t Z\"{u}rich, Winterthurerstrasse 190, CH-8057 Z\"{u}rich, Switzerland}

\author{O.~K.~Forslund}
\affiliation{Physik-Institut, Universit\"{a}t Z\"{u}rich, Winterthurerstrasse 190, CH-8057 Z\"{u}rich, Switzerland}
\affiliation{Department of Physics and Astronomy, Uppsala University, Box 516, SE-75120 Uppsala, Sweden}

\author{J.~Lass}
\affiliation{PSI Center for Neutron and Muon Sciences, Paul Scherrer Institut, CH-5232 Villigen-PSI, Switzerland}
\affiliation{Laboratory for Quantum Magnetism, Institute of Physics, Ecole Polytechnique F\'ed\'erale de Lausanne (EPFL), CH-1015 Lausanne, Switzerland}

\author{T.~Shiroka}
\affiliation{PSI Center for Neutron and Muon Sciences, Paul Scherrer Institut, CH-5232 Villigen-PSI, Switzerland}

\author{A.~Linden}
\affiliation{Department of Chemistry, University of Zurich, Winterthurerstrasse 190, CH-8057 Z\"{u}rich, Switzerland}

\author{D.~G.~Mazzone}
\affiliation{PSI Center for Neutron and Muon Sciences, Paul Scherrer Institut, CH-5232 Villigen-PSI, Switzerland}

\author{J.~Ollivier}
\affiliation{Institut Laue Langevin, BP156, 38042 Grenoble, France}

\author{S.~Shin}
\affiliation{PSI Center for Neutron and Muon Sciences, Paul Scherrer Institut, CH-5232 Villigen-PSI, Switzerland}

\author{M.~Medarde}
\affiliation{PSI Center for Neutron and Muon Sciences, Paul Scherrer Institut, CH-5232 Villigen-PSI, Switzerland}

\author{B.~Lake}
\affiliation{Helmholtz-Zentrum Berlin f\"{u}r Materialien und Energie, Hahn-Meitner-Platz 1, 14109 Berlin, Germany}
\affiliation{Institut f\"{u}r Festk\"{o}rperforschung, Technische Universit\"{a}t Berlin, 10623 Berlin, Germany}

\author{M.~M\aa nsson}
\affiliation{Department of Applied Physics, KTH Royal Institute of Technology, SE-106 91 Stockholm, Sweden}

\author{M.~Bartkowiak}
\affiliation{PSI Center for Neutron and Muon Sciences, Paul Scherrer Institut, CH-5232 Villigen-PSI, Switzerland}

\author{B.~Normand}
\affiliation{PSI Center for Scientific Computing, Theory and Data, Paul Scherrer Institut, CH-5232 Villigen-PSI, Switzerland}
\affiliation{Laboratory for Quantum Magnetism, Institute of Physics, Ecole Polytechnique F\'ed\'erale de Lausanne (EPFL), CH-1015 Lausanne, Switzerland}

\author{P.~K\"{o}gerler}
\affiliation{Institute of Inorganic Chemistry, RWTH Aachen University, 52056 Aachen, Germany}

\author{Y.~Sassa}
\affiliation{Department of Physics, Chalmers University of Technology, SE-41296 G\"{o}teborg, Sweden}

\author{M.~Janoschek}
\affiliation{PSI Center for Neutron and Muon Sciences, Paul Scherrer Institut, CH-5232 Villigen-PSI, Switzerland}
\affiliation{Physik-Institut, Universit\"{a}t Z\"{u}rich, Winterthurerstrasse 190, CH-8057 Z\"{u}rich, Switzerland}

\author{G.~Simutis}
\email{gediminas.simutis@psi.ch}
\affiliation{PSI Center for Neutron and Muon Sciences, Paul Scherrer Institut, CH-5232 Villigen-PSI, Switzerland}
\affiliation{Department of Physics, Chalmers University of Technology, SE-41296 G\"{o}teborg, Sweden}

\date{\today}

\begin{abstract}

Low-dimensional quantum magnets are a versatile materials platform for studying the emergent many-body physics and collective excitations that can arise even in systems with only short-range interactions. Understanding their low-temperature structure and spin Hamiltonian is key to explaining their magnetic properties, including unconventional quantum phases, phase transitions, and excited states. We study the metal–organic coordination compound (C$_5$H$_9$NH$_3$)$_2$CuBr$_4$ and its deuterated counterpart, which upon its discovery was identified as a candidate two-leg quantum ($S = 1/2$) spin ladder in the strong-leg coupling regime. By growing large single crystals and probing them with both bulk and microscopic techniques, we deduce that two previously unknown structural phase transitions take place between 136~K and 113~K. The low-temperature structure has a monoclinic unit cell that gives rise to two inequivalent spin ladders. We further confirm the absence of long-range magnetic order down to 30~mK and investigate the implications of this two-ladder structure for the magnetic properties of (C$_5$H$_9$NH$_3$)$_2$CuBr$_4$ by analyzing our own specific-heat and susceptibility data.

\end{abstract}

\maketitle{}

\section{Introduction\label{sec:intro}}

One-dimensional quantum magnets provide a testbed for many-body quantum physics, because experimental measurements of their intrinsically collective excitations can be described by powerful analytical and numerical techniques. One particularly versatile model system is the two-leg $S = 1/2$ quantum spin ladder, which with isotropic (Heisenberg) interactions is described by only two parameters, $J_\mathrm{leg}$ for the ladder legs and $J_\mathrm{rung}$ for the rungs, and hence by a single ratio, $\alpha = J_\mathrm{leg}/J_\mathrm{rung}$ \cite{Dagotto1996}. Although these ladders have a spin gap for any finite $\alpha$ and can all be described in a resonating valence-bond framework with different correlation distributions \cite{White1994}, the zero-field spectral function varies widely from a single triplon branch in strong-rung ladders ($\alpha < 1/2$) to weakly confined spinons in the spin-chain limit ($\alpha \gg 1$) \cite{Schmidiger2013}. 

Arguably the most interesting properties of two-leg ladders appear in an applied magnetic field strong enough to close the spin gap, where the system becomes a spin Tomonaga–Luttinger liquid (TLL), a theoretical model describing interacting fermions in one dimension \cite{Giamarchi1999}. Early ladder materials, based on cuprate perovskites, included SrCu$_2$O$_3$ \cite{hiroi1991,azuma1994,ishida1994,kojima1995}, LaCuO$_{2.5}$ \cite{hiroi1995}, and (Sr$_{14-x}$Ca$_x$)Cu$_{24}$O$_{41}$ \cite{mayaffre1998,eccleston1998}; these systems had $\alpha \approx 1$ with $J_\mathrm{leg}$ and $J_\mathrm{rung}$ both very large, and thus far outside the range of laboratory magnetic fields. Metal-organic materials based on Cu$^{2+}$ ions offered a solution to producing low-$J$ ladders, 
beginning with the candidate ladder compound (C$_5$H$_{12}$N$_2$)$_2$Cu$_2$Cl$_4$ \cite{Chaboussant_1997,Chaboussant_1998_NMR},
and much of the TLL phenomenology was discovered using the strong-rung system (C$_5$H$_{12}$N)$_2$CuBr$_4$ (BPCB) \cite{Klanjsek2008,Ruegg2008}. This included triplon fractionalization \cite{thielemann2009a}, three-dimensional ordering \cite{Thielemann2009}, and the full spectral function of all three field-split triplon branches \cite{bouillot2011}. Two-leg ladders have also been used as a platform for observing two-triplon bound states in the absence of frustration \cite{windt01,Schmidiger2012,ward2017} and, in the presence of strong frustration, for the theoretical study of fully localized quasiparticles, exact bound states, and anomalous thermodynamics \cite{Xian1995,Honecker2016,Honecker2016a}.

Materials in the strong-leg regime, {$\alpha > 1$}, nevertheless retain a special interest due to the delocalized and spinonic character of their correlations and excitations. To date \chemDIMPY\ (DIMPY) is the only clean, strong-leg ladder compound to be studied in detail, with extensive bulk and spectroscopic studies performed to unravel its magnetic properties \cite{shapiro_synthesis_2007,Hong2010,schmidiger_long-lived_2011,Schmidiger2012,Schmidiger2013,Schmidiger2013M}. In the TLL, it was shown that the interaction between the emergent fermions depends both on $\alpha$ and on the applied field, such that in DIMPY it could be controlled and made attractive by increasing the field \cite{Ninios2012,jeong_attractive_2013, Povarov2015, jeong_dichotomy_2016}. In the direction of controlled disorder physics, it was found that, when depleted by 
nonmagnetic impurities, both DIMPY and BPCB show a universal $LT$ scaling of their staggered susceptibilities \cite{Galeski_2022_QSL}
and that the ladders in DIMPY host emergent strongly interacting spin islands \cite{schmidiger_emergent_2016}.

\begin{figure*}[t]
    \begin{subfigure}%
        \centering
        \includegraphics[width={0.49\textwidth}]{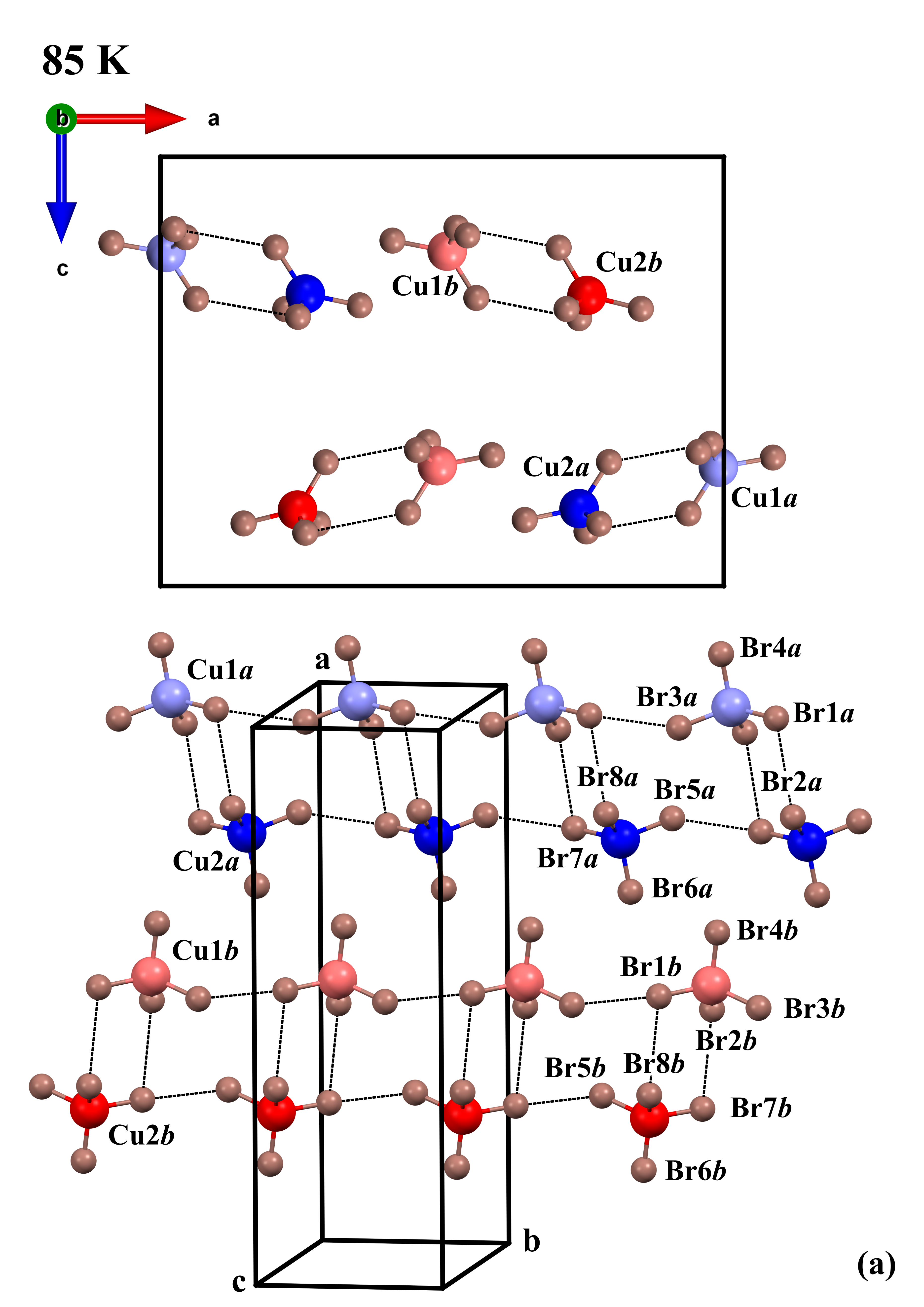}
    \end{subfigure}
    \begin{subfigure}
        \centering
        \includegraphics[width={0.49\textwidth}]{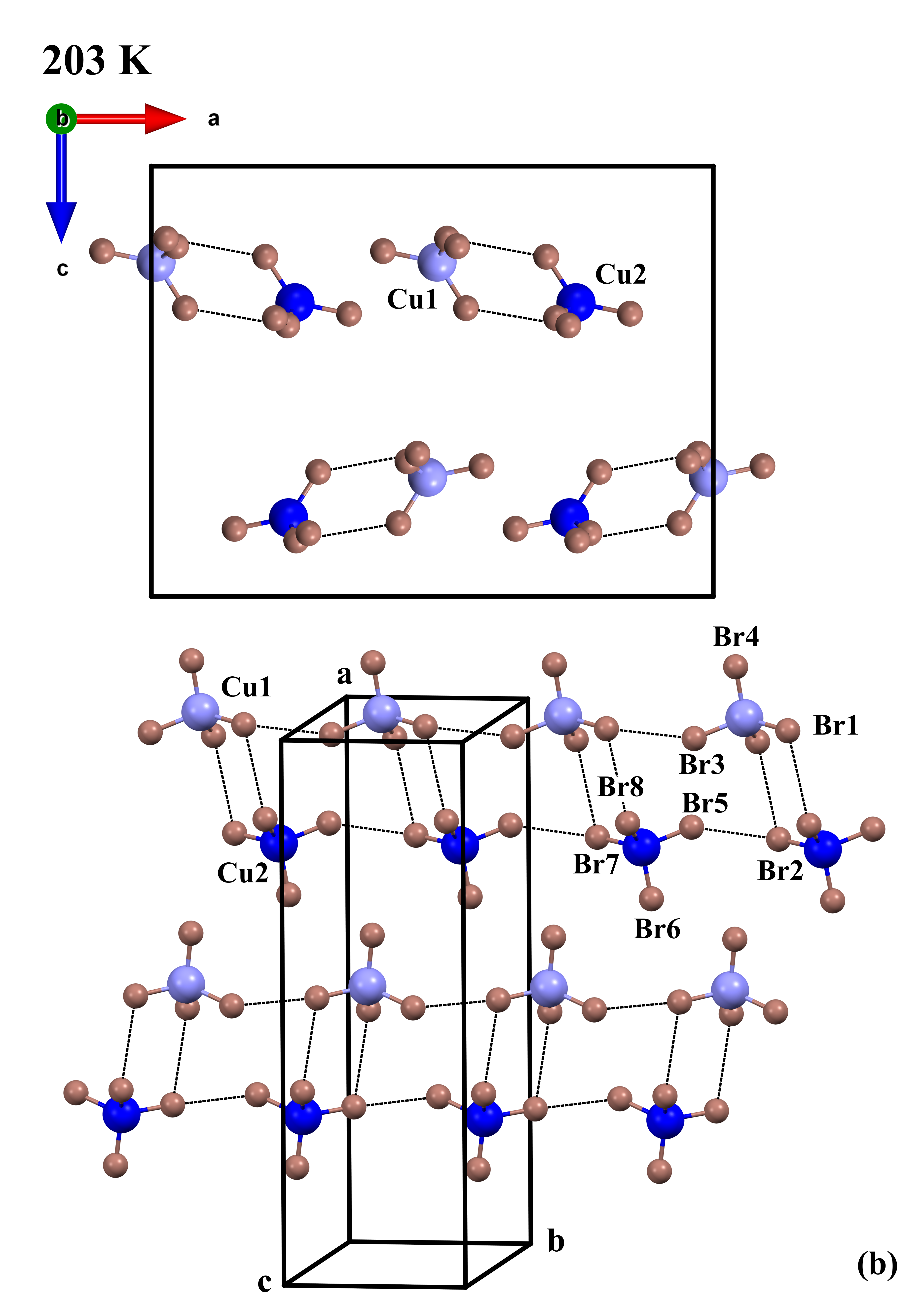}
    \end{subfigure}
    \caption{\label{fig:structure}%
Key structural elements of \CPA\ measured at $T = 85$ (a) and 203~K (b). The upper panels show the ladder rungs (dimers) viewed down the $b$ axis. The lower panels provide a perspective view of two of the four ladders in the unit cell. The shortest paths connecting Br$^-$ ions are shown as black lines. The Cu$^{2+}$ ions on opposite sides of every ladder rung, marked Cu1 and Cu2, are inequivalent at all temperatures, and below the structural phase transitions there are four inequivalent copper sites (shown as light and dark blue and light and dark red) forming two inequivalent ladders. Because the magnetic interactions depend sensitively on the Cu--Br$\cdots$Br--Cu geometry, we report the different interatomic distances and angles for the 85~K structure in Table~\ref{tab:cpa_contacts}.} 
\end{figure*}

Despite this level of understanding, DIMPY has also been found to exhibit field-induced low-temperature phases that are not expected for the ideal two-leg ladder \cite{Jeong2017}, pointing to the presence of additional terms in the spin Hamiltonian. Theoretical studies have shown that readily anticipated extra terms, such as Dzyaloshinskii-Moriya (DM) interactions and frustration, have substantial effects on the properties and phase diagram \cite{Penc2007,Michaud2010,Weichselbaum2010}. 
Experimental studies of BPCB found unexpected dynamics in the TLL regime, an orientation-dependent spin gap, and anomalous $g$-factor values \cite{Krasnikova_2020}. In DIMPY, electron spin-resonance (ESR) measurements found gapped modes with an unconventional, nonlinear frequency-field dependence \cite{Ozerov_2015_ESR} and line broadening \cite{Glazkov_2015_ESR} that were related to DM interactions. 
For this reason, additional materials of the strong-leg ladder type are required to separate universal from non-universal properties. Further, given the change in ladder nature as a function of $\alpha$, additional materials spanning the full $\alpha$ range are required for detailed experimental analysis of the crossover from triplonic to spinonic physics in quantum spin ladders. 

To expand the platform of model materials, in this paper we investigate \chemCPA\ and its deuterated counterpart, \chemCPAD, to both of which we refer as Cu-CPA. Upon its discovery, this compound was proposed as a candidate strong-leg spin ladder with $\alpha = 2.11$ \cite{Willett2004}. At room temperature, \CPA\ has an orthorhombic structure in which the Cu$^{2+}$ ions are linked by halide (Br$\cdots$Br) bonds to form a structure of well-isolated magnetic motifs separated by organic cations. At $T_{2a} = 260$~K, the authors of Ref.~\cite{Willett2004} found a structural phase transition accompanied by a doubling of the crystallographic $a$ axis, with the unit cell remaining orthorhombic.

The bond lengths and magnetic interaction pathways suggest that this phase of \CPA, which is shown in Fig.~\ref{fig:structure}(b), should realize a strong-leg ladder, with any further-neighbor interactions, including diagonal and interladder pathways, being negligible. Initial magnetic susceptibility measurements support this scenario, but are far from conclusive. More detailed studies of \CPA\ have, however, been hampered by the difficulty in producing sizeable single crystals. Here we overcome this challenge by optimizing the crystal growth from solution and thus obtaining large single crystals of both hydrogenated and deuterated \CPA.

We have performed specific-heat, susceptibility, and detailed structural measurements on \CPA, which taken together reveal that the low-temperature structure is significantly richer than the initial studies suggested. In particular, the system undergoes two more structural phase transitions below $T_{2a}$, which take it into a low-temperature monoclinic phase. Using neutron and X-ray diffraction, we establish that at low temperatures \CPA\ contains two structurally inequivalent ladders, as shown in Fig.~\ref{fig:structure}(a).

This paper is organized as follows. In Sec.~\ref{sec:experiments} we describe the materials and methods used in our study. We present our experimental results for the presence of two structurally inequivalent ladders in Sec.~\ref{sec:results}. In Sec.~\ref{sec:magint} we turn to the magnetic interactions, discussing the qualitative consequences of the two-ladder structure for the spin Hamiltonian of \CPA.
In Sec.~\ref{sec:magprop} we extend this investigation to the magnetic properties, presenting our low-temperature specific-heat data to obtain a more quantitative analysis of the observable consequences of the two-ladder nature.
A brief conclusion is provided in Sec.~\ref{sec:conclusion}. 

\section{Materials and methods\label{sec:experiments}}

\subsection{Crystal Growth} 

The single crystals of \CPA\ used for this study were synthesized using growth from solution. The synthesis method reported earlier~\cite{Willett2004} was optimized to produce large single crystals. Deuterated versions of the compound were produced in the same manner, in order to make possible high-resolution neutron scattering experiments. 

We focus our description on the synthesis of \chemCPAD. First, a 47 weight \% DBr solution in D$_2$O (18.20~ml, 0.129~mol) was added dropwise to a solution of cyclopentylamine-$d_{11}$ (11.25~g, 0.117~mol) in 20~ml of D$_2$O. The resulting mixture was stirred for five minutes and left to stand for slow evaporation in the fume hood until white crystals appeared. Crystals of cyclopentylammonium bromide-$d_{12}$ (CPA-DBr) were filtered and dried in vacuum for five hours. Subsequently, CPA-DBr (0.88~g, 5~mmol) was dissolved in 6~ml of D$_2$O and, to this mixture, a solution of CuBr$_2$ (0.56~g, 2.50~mmol) in 10~ml of D$_2$O was added dropwise and stirred for five minutes. To this final solution, 2~ml of 48\% DBr in D$_2$O were added dropwise to avoid hydrolysis [formation of Cu(OD)$_2$]. The solution was filtered and left to stand in a beaker for slow evaporation. After three months, long black needles (of approximate size $20 \times 2 \times 2$~mm) grew as single crystals in the mother solution.

\subsection{Specific Heat}

The specific heat was measured at zero field (ZF) and in an applied magnetic field of $\mu_0H = 7$~T in a Quantum Design Physical Property Measurement System (PPMS) for the respective temperature ranges 4--200~K and 80--160~K. Low-temperature measurements over the range 0.36--20~K were performed, using a Quantum Design $^3$He insert for the PPMS, to provide sufficient overlap with the conventional $^4$He measurements. The standard ZF measurements were performed on a twinned, deuterated \CPA\ crystal [mass 7.00(1)~mg] and repeated with a single, deuterated crystal of mass 1.34(1)~mg. The measurements in field were performed on the same single crystal [mass 1.34(1)~mg] used for the standard ZF measurements. The ${}^{3}$He measurements were performed on a different deuterated single crystal, also of mass 1.34(1)~mg. Finally, the ZF measurements (over temperature range 4--200~K) on hydrogenated \CPA\ were performed on a single crystal of mass 1.10(1)~mg. 

\subsection{X-ray diffraction}

Single-crystal X-ray diffraction measurements were performed with a Stadivari diffractometer (STOE) equipped with a liquid-nitrogen open-flow cooler (Oxford Cryosystems, Cryostream) that enabled the acquisition of X-ray diffraction data down to 85~K. Monochromated Mo K$\alpha$ radiation was used and full structural datasets were acquired at 85, 125, and 203~K, while further, partial datasets were acquired at 95, 105, 115, and 150~K. High-resolution X-ray powder diffraction measurements were performed on the MS beamline \cite{Willmott2013} at the Swiss Light Source (PSI) on capillary samples using the Mythen~III detector. A wavelength of 0.99952~\AA, as calibrated with a silicon standard from NIST (SRM 640d), was used for these measurements, while the temperature was controlled using an Oxford Cryosystems Cryostream.

\subsection{Neutron scattering}

Additional neutron scattering experiments were performed to confirm the absence of further structural or magnetic phase transitions down to millikelvin temperatures. These used the multiplexing spectrometer CAMEA at the Swiss Spallation Neutron Source (SINQ, PSI) \cite{Groitl2016_camea,Lass2023} in order to reduce the inelastic background from the sample and sample holder while searching for possible weak magnetic Bragg peaks. The neutron experiments were conducted in a dilution refrigerator attaining temperatures of 30~mK, and the data analyzed with the software package MJOLNIR \cite{Lass2020}.

\section{Crystal structure and phase transitions\label{sec:results}}

\subsection{Specific heat\label{ssec:specific_heat}}

%==== figure =============================%
\begin{figure}[t]
\includegraphics[width={\columnwidth}]{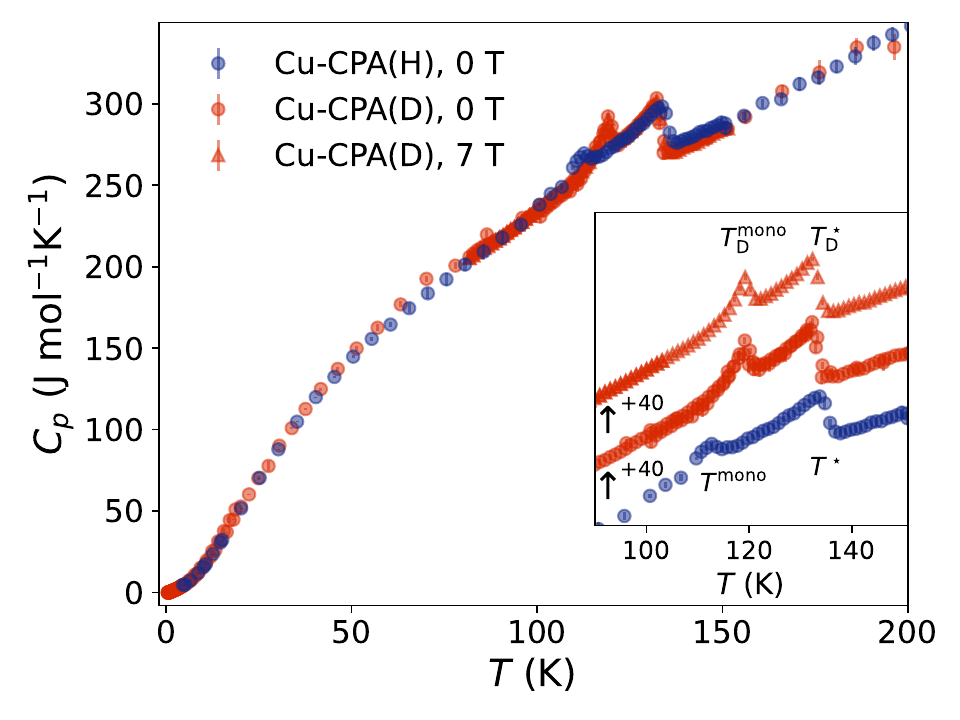}
\caption{\label{fig:heatcapacity}%
Specific heat ($C_p$) as a function of temperature ($T$), measured for deuterated \CPA\ samples at ZF (red, circles) and under a magnetic field of 7~T (red, triangles), and for one hydrogenated \CPA\ sample at ZF (blue, circles). The two peaks at 119 and 132~K for deuterated \CPA, and at 113 and 136~K for hydrogenated \CPA, indicate previously unreported structural phase transitions. Inset: detail of the two phase transitions; the data are displayed with a relative vertical offset. $T^\mathrm{mono}$ labels the orthorhombic-to-monoclinic transition and $T^\star$ an ordering of the organic groups.}
\end{figure}

The specific-heat data shown in Fig.~\ref{fig:heatcapacity} were obtained from three different measurements covering a temperature range (below 200~K) not studied previously. We observe two distinct peaks occurring at temperatures we label as $T^{\star} = 136$~K and $T^\mathrm{mono} = 113$~K in hydrogenated \CPA, with the corresponding peaks for deuterated \CPA\ appearing over a slightly narrower range. We defer to Secs.~\ref{ssec:monoclinic} and \ref{ssec:order-disorder} the explanation of how these two peaks are related to two structural phase transitions.

This observation is unexpected, as to date it had been assumed that the low-temperature structure is achieved below the structural phase transition measured at $T_{2a} = 260$~K \cite{Willett2004,Woodward2005}. Thus we performed multiple heating and cooling cycles on both the hydrogenated and deuterated compounds in order to confirm that both phase transitions are reversible, reproducible, and independent of the measurement history. Neither phase transition is affected by magnetic fields up to 7~T, further reinforcing the deduction that both are of structural nature.

Although the transitions are sample-independent, they do exhibit an isotope effect (Fig.~\ref{fig:heatcapacity}, inset). Compared to the hydrogenated version of the crystals, the specific-heat peaks in the deuterated samples appear at the slightly different temperatures ${T_\mathrm{D}^\star = 132}$~K and $T_\mathrm{D}^\mathrm{mono} = 119$~K. Such a change in transition temperatures is a common occurrence in metal-organic systems \cite{Wilson2015, Shi2018}, arising due to the change in donor-acceptor distance within the hydrogen bonds. 

Below 100~K, the specific heat varies smoothly down to our lowest measured temperature of 360~mK (Fig.~\ref{fig:heatcapacity}). This is consistent with the magnetically disordered ground state expected in a two-leg quantum spin ladder, which is also suggested by measurements of the magnetic susceptibility down to 2~K performed in Ref.~\cite{Willett2004}. Our neutron diffraction measurements further confirmed the absence of any magnetic Bragg peaks down to 30~mK.

%==== figure =============================%
\begin{figure}[t]
\centering
\includegraphics[width={\columnwidth}]{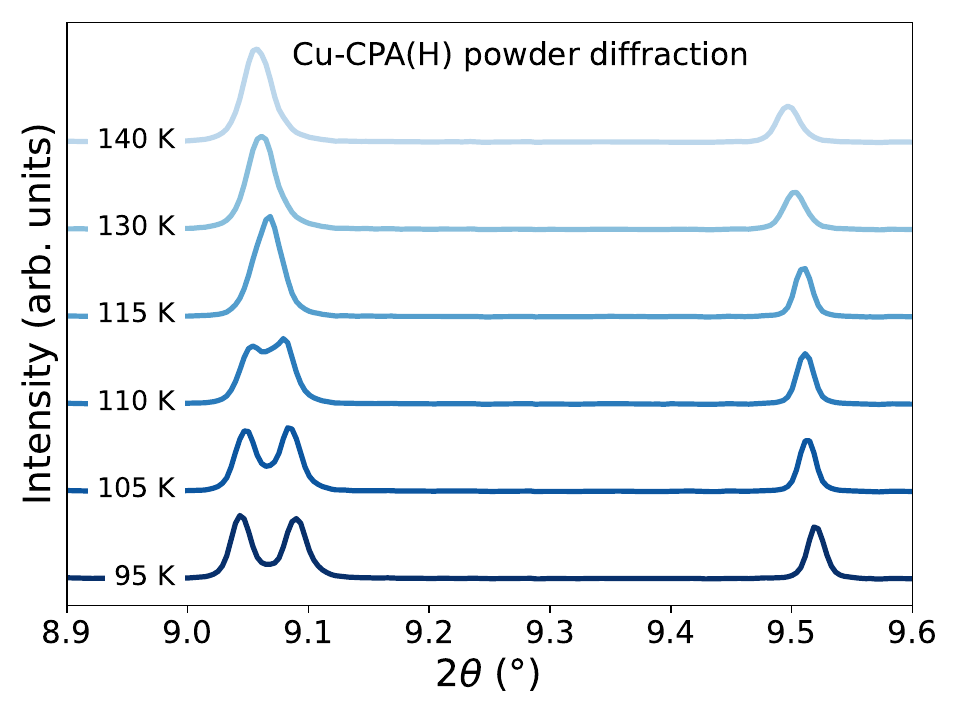}
\caption{\label{fig:powderXRD}
X-ray powder diffraction patterns obtained for hydrogenated \CPA\ on lowering the temperature. The splitting of the peak at $2\theta = 9.08$° between 115 and 110~K indicates a lifting of the degeneracy between the Bragg peaks $(2,1,-1)$ and $(2,1,1)$. This is a signature of the transition into a monoclinic crystal structure, consistent with $T^\mathrm{mono}$ in Fig.~\ref{fig:heatcapacity}. The peak at $2\theta = 9.51$°, corresponding to the Bragg peaks $(\pm4, 0, 0)$, remain degenerate.}
\end{figure}

\subsection{Crystal Structure}

We analyzed the crystal structure of \CPA\ through the two phase transitions by thorough X-ray and neutron scattering experiments. First we confirmed the previously known phase transition at $T_{2a} = 260$~K and refined the crystal structure at 203~K, following Ref.~\cite{Willett2004}. This structure has orthorhombic space group $Pna2_1$ with lattice parameters $a = 23.9927(6)$~\AA, $b = 8.0894(6)$~\AA, $c = 18.3449(6)$~\AA, and is shown in Fig.~\ref{fig:structure}(b). We adopt it as a frame of reference for the remainder of our discussion. We then analyzed the crystal structure of \CPA\ at 85~K, finding the results displayed in Fig.~\ref{fig:structure}(a) and summarized in App.~\ref{sec:structural_properties}. We now concentrate on the low-temperature regime ($T \le 140$~K) in order to relate the specific-heat peaks to two structural phase transitions.

%==== figure =============================%
\begin{figure}[t]
\centering
\includegraphics[width={\columnwidth}]{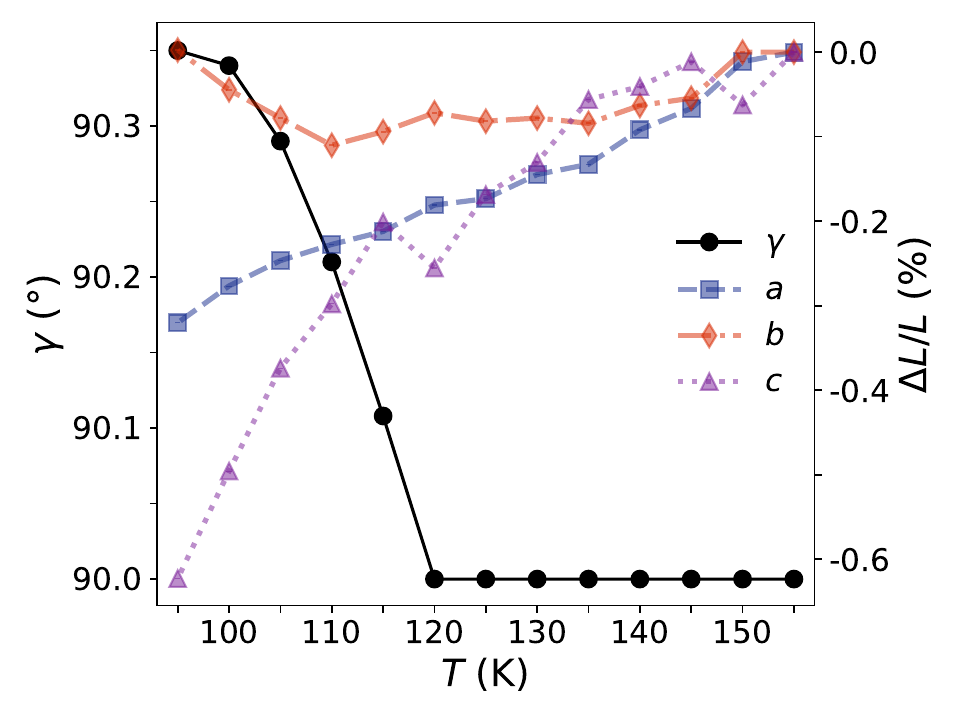}
\caption{\label{fig:thermal_evolution} 
Thermal evolution of the structural parameters of hydrogenated \CPA. Relative changes to all three lattice parameters ($\Delta L/L$) are indexed to the right axis and the monoclinic angle ($\gamma$) to the left axis. All lines serve only to join the data points for visual clarity.
}
\end{figure}

\subsubsection{Monoclinic transition \label{ssec:monoclinic}}

Figure \ref{fig:powderXRD} shows temperature-dependent data from powder X-ray diffraction performed on hydrogenated \CPA. At high temperatures, the peaks at $2\theta = 9.08$° and $2\theta = 9.51$° correspond respectively to the Bragg peaks [$(2,1,-1)$, $(2,1,1)$] and $(4,0,0)$, as defined in the orthorhombic crystal structure of the system at 203~K. The splitting of the peak at $2\theta = 9.08$° below 115~K corresponds to a structural phase transition and a decrease of crystal symmetry from orthorhombic to monoclinic. On passing through this transition, the crystallographic angle $\gamma$ changes from 90° to 90.35°, which is manifested most clearly in a splitting of mixed Bragg peaks that involve the $c$ direction, such as $(2,1,-1)$ and $(2,1,1)$. Above 115~K, these peaks coincide by symmetry, but the structural phase transition leads to a change of space group from $Pna2_1$ to $P112_1$ and a lifting of degeneracies. Because the onset temperature of the splitting is consistent with the lower-temperature peak observed in the specific heat, we assign this phase transition to $T^\mathrm{mono}$. 

It is this orthorhombic-to-monoclinic transition that causes the ladders to become pairwise structurally inequivalent [Fig.~\ref{fig:structure}(a)]. 
The temperature-dependence of the lattice parameters $a$, $b$, and $c$, as well as of $\gamma$, through $T^\mathrm{mono}$ are shown in Fig.~\ref{fig:thermal_evolution} (the two angles $\alpha$ and $\beta$ remain 90.00°). The change of the monoclinic angle indicates that this phase transition is a continuous process, saturating around $\gamma \simeq 90.35$°.

\subsubsection{Order-disorder transition \label{ssec:order-disorder}}

Turning to the phase transition at $T^\star$, our structural measurements revealed no additional lowering of symmetry at this temperature. Instead we ascribe the atomic reorganization taking place at $T^\star$ to an order-disorder transition, motivated by the demonstration in Ref.~\cite{Willett2004} of disorder among the organic cations. Specifically, below the structural transition at $T_{2a}$, the 5 carbon atoms within one of the four cyclopentylammonium groups in the unit cell (denoted as C16A--C20A) can adopt a second position (C16B--C20B), which appears with a probability of 45\% \cite{Willett2004}. To investigate this situation, we collected a full structural dataset by X-ray diffraction on a deuterated single crystal of \CPA\ at 125~K (i.e.~directly below $T^\star$, but still above $T^\mathrm{mono}$). In contrast to the 203~K dataset from the same crystal, the refinement at 125~K did not require the inclusion of any such disorder in the C16--C20 atoms. This result confirms that the cyclopentylammonium groups become fully ordered below $T^\star$ and hence that this transition is of order-disorder type.

\begin{figure}[t]
\centering
\includegraphics[width={\columnwidth}]{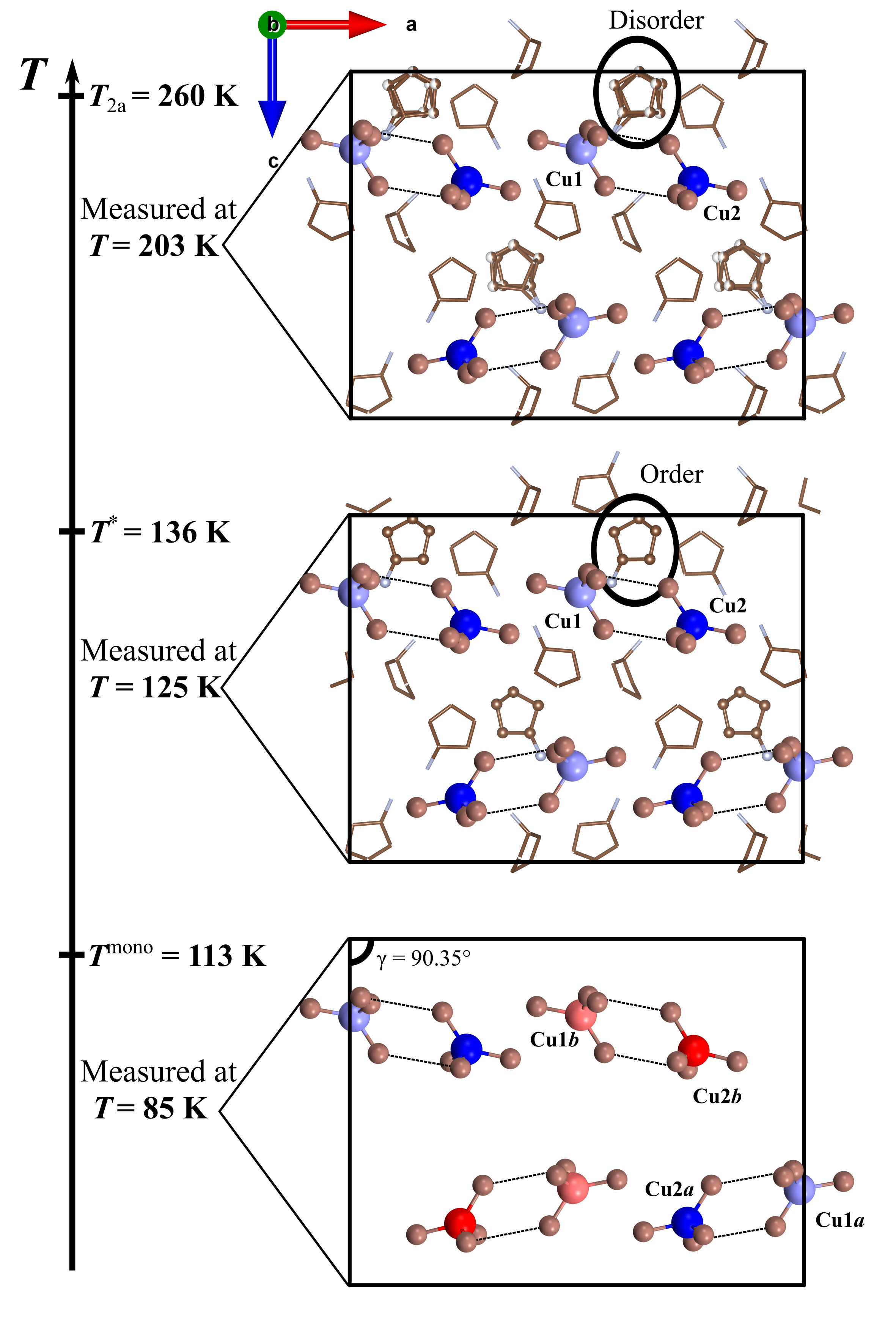}
\caption{\label{fig:summary}
Summary of the crystal structures and phase transitions revealed by our measurements on hydrogenated \CPA. Phase transitions and our naming convention are marked in bold text. At temperatures from $T_{2a}$ down to $T^*$ is a structure with disorder in one of the four organic cation groups (upper panel), as reported in Ref.~\cite{Willett2004} and measured at 203~K. The positions of the disordered carbon atoms are represented as the double fivefold rings of small brown-white spheres. At temperatures from $T^*$ down to $T^{\rm mono}$, these organic cations have become ordered (center panel), as measured at 125~K. At temperatures below $T^{\rm mono}$, the system adopts the monoclinic structure, with a continuous transition of the monoclinic angle to a low-temperature value of $\gamma = 90.35$° (Fig.~\ref{fig:thermal_evolution}). In the representation of the structure measured at 85~K (lower panel), the cyclopentylammonium groups are omitted for clarity.
}

\end{figure}

\subsubsection{Summary of phase transitions \label{ssec:lattice_evolution}}

To clarify the complex structural evolution of \CPA, we summarize the situation in Fig.~\ref{fig:summary}. Proceeding from high to low temperatures, the one-ladder structure is found below $T_{2a} = 260$~K \cite{Willett2004}, in a structure with four distinct cyclopentylammonium groups of which one shows two different configurations with an approximately 50:50 distribution. In the upper panel of Fig.~\ref{fig:summary}, the non-disordered groups are shown only as spokes, the disordered group as the doubled ball-and-spoke structure. Below $T^* = 136$~K (${T_\mathrm{D}^\star = 132}$~K), these groups select only one of the two configurations, thereby lifting the disorder, with no other discernible change in structure (center panel). Only below $T^\mathrm{mono} = 113$~K ($T_\mathrm{D}^\mathrm{mono} = 119$~K) does the system adopt its low-temperature, monoclinic structure, where the two ladder units become structurally inequivalent (lower panel). 

\section{Magnetic pathways and magnetic interactions\label{sec:magint}}

Interpretation of the phenomena observed in quantum magnetic materials depends crucially on the availability of realistic model Hamiltonians. For this the underlying crystal structure provides essential insight into the symmetry, number, and relative strengths of the relevant interaction parameters. For Cu-CPA we have found that the two-parameter, strong-leg ladder Hamiltonian assumed previously is in fact incomplete. The discovery of two inequivalent ladders at low temperatures requires that existing measurements be interpreted in a different light and should establish the foundation for future spectroscopic studies of this model two-ladder material.

\begin{table}
	\centering
	\caption{Cu--Br bond lengths [\AA] and Br--Cu--Br angles [deg] in the \CPA\ structure at 85~K and at 203~K. The atomic site notation is that of Fig.~\ref{fig:structure}.
	\label{tab:cpa_CuBr_distances}}
	\begin{tabular}{@{}l l l@{}}
		\toprule
        $\;\;$ Bonds & $T = 85$~K & $T = 203$~K \\
    %   $\;\;$ Bonds & $d_{85~\mathrm{K}}$ [\AA] & & & Bonds & $d_{85~\mathrm{K}}$ [\AA] & & $d_{203~\mathrm{K}}$ [\AA]\\
        \midrule
		$\;\;$ Cu$_{1a}$--Br$_{1a}$ & 2.368(4) & \multirow{2}{*}{2.358(2)}\\
        $\;\;$ Cu$_{1b}$--Br$_{1b}$ & 2.367(4) \\
	    $\;\;$ Cu$_{1a}$--Br$_{2a}$ & 2.370(3) & \multirow{2}{*}{2.359(2)}\\
        $\;\;$ Cu$_{1b}$--Br$_{2b}$ & 2.370(4)\\
        $\;\;$ Cu$_{1a}$--Br$_{3a}$ & 2.399(3) & \multirow{2}{*}{2.398(2)}\\
        $\;\;$ Cu$_{1b}$--Br$_{3b}$ & 2.398(3)\\
        $\;\;$ Cu$_{1a}$--Br$_{4a}$ & 2.388(3) & \multirow{2}{*}{2.378(2)}\\
        $\;\;$ Cu$_{1b}$--Br$_{4b}$ & 2.384(4)\\
        \\
        $\;\;$ Cu$_{2a}$--Br$_{5a}$ & 2.404(3) & \multirow{2}{*}{2.388(2)}\\
        $\;\;$ Cu$_{2b}$--Br$_{5b}$ & 2.399(3)\\
	    $\;\;$ Cu$_{2a}$--Br$_{6a}$ & 2.392(3) & \multirow{2}{*}{2.389(2)}\\
        $\;\;$ Cu$_{2b}$--Br$_{6b}$ & 2.394(3)\\
        $\;\;$ Cu$_{2a}$--Br$_{7a}$ & 2.365(3) & \multirow{2}{*}{2.360(2)}\\
        $\;\;$ Cu$_{2b}$--Br$_{7b}$ & 2.363(3)\\
        $\;\;$ Cu$_{2a}$--Br$_{8a}$ & 2.376(4) & \multirow{2}{*}{2.368(2)}\\
        $\;\;$ Cu$_{2b}$--Br$_{8b}$ & 2.380(3)\\
        \\
        $\;\;$ Br$_{1a}$--Cu$_{1a}$--Br$_{2a}$ & 99.54(12) & \multirow{2}{*}{99.06(8)}\\
        $\;\;$ Br$_{1b}$--Cu$_{1b}$--Br$_{2b}$ & 99.18(13)\\
        $\;\;$ Br$_{1a}$--Cu$_{1a}$--Br$_{3a}$ & 129.67(12) & \multirow{2}{*}{127.50(10)}\\
        $\;\;$ Br$_{1b}$--Cu$_{1b}$--Br$_{3b}$ & 128.05(14)\\
        $\;\;$ Br$_{1a}$--Cu$_{1a}$--Br$_{4a}$ & 101.20(12) & \multirow{2}{*}{102.08(9)}\\
        $\;\;$ Br$_{1b}$--Cu$_{1b}$--Br$_{4b}$ & 102.13(15)\\
        $\;\;$ Br$_{2a}$--Cu$_{1a}$--Br$_{3a}$ & 97.87(11) & \multirow{2}{*}{98.45(8)}\\
        $\;\;$ Br$_{2b}$--Cu$_{1b}$--Br$_{3b}$ & 99.13(12)\\
        $\;\;$ Br$_{2a}$--Cu$_{1a}$--Br$_{4a}$ & 134.10(12) & \multirow{2}{*}{133.61(10)}\\
        $\;\;$ Br$_{2b}$--Cu$_{1b}$--Br$_{4b}$ & 132.15(14)\\
        $\;\;$ Br$_{3a}$--Cu$_{1a}$--Br$_{4a}$ & 99.52(12) & \multirow{2}{*}{100.48(8)}\\
        $\;\;$ Br$_{3b}$--Cu$_{1b}$--Br$_{4b}$ & 100.44(13)\\
        \\
        $\;\;$ Br$_{5a}$--Cu$_{2a}$--Br$_{6a}$ & 98.50(12) & \multirow{2}{*}{99.23(7)}\\
        $\;\;$ Br$_{5b}$--Cu$_{2b}$--Br$_{6b}$ & 99.10(12)\\
        $\;\;$ Br$_{5a}$--Cu$_{2a}$--Br$_{7a}$ & 130.60(12) & \multirow{2}{*}{129.75(9)}\\
        $\;\;$ Br$_{5b}$--Cu$_{2b}$--Br$_{7b}$ & 131.28(12)\\
        $\;\;$ Br$_{5a}$--Cu$_{2a}$--Br$_{8a}$ & 98.48(12) & \multirow{2}{*}{98.62(7)}\\
        $\;\;$ Br$_{5b}$--Cu$_{2b}$--Br$_{8b}$ & 97.84(12)\\
        $\;\;$ Br$_{6a}$--Cu$_{2a}$--Br$_{7a}$ & 100.31(12) & \multirow{2}{*}{100.87(8)}\\
        $\;\;$ Br$_{6b}$--Cu$_{2b}$--Br$_{7b}$ & 99.98(12)\\
        $\;\;$ Br$_{6a}$--Cu$_{2a}$--Br$_{8a}$ & 135.72(13) & \multirow{2}{*}{134.83(10)} $\;\;$\\
        $\;\;$ Br$_{6b}$--Cu$_{2b}$--Br$_{8b}$ & 134.66(12)\\
        $\;\;$ Br$_{7a}$--Cu$_{2a}$--Br$_{8a}$ $\;\;\;\;\;\;\;\;$ & 98.94(12) $\;\;\;\;\;\;\;\;$ & \multirow{2}{*}{98.81(7)}\\
        $\;\;$ Br$_{7b}$--Cu$_{2b}$--Br$_{8b}$ & 99.65(12)\\
    \bottomrule
    \end{tabular}
\end{table}

\subsection{Pathways between magnetic ions \label{ssec:pathways}}

The magnetic interactions in insulating materials result from extended superexchange processes on the pathways between magnetic ions. Because they depend on the overlap of electronic orbitals along these pathways, they have a strong and highly nonlinear dependence on the interatomic separations (``bond lengths'') and on the angles between bonds \cite{Turnbull2005,Foyevtsova2011}. To tabulate all the information required to estimate the relevant magnetic interactions, in Table~\ref{tab:cpa_CuBr_distances} we first report the {Cu--Br} bond lengths and angles within the CuBr$_{4}^{2-}$ anions. The primary differences between the 85~K structure and the 203~K structure reported previously are a consequence of the orthorhombic-to-monoclinic transition, which leads to a minor deformation of the CuBr$_4$ tetrahedra [Fig.~\ref{fig:structure}(a)]. This deformation creates four inequivalent Cu sites in the low-temperature structure, compared to two above $T^\mathrm{mono}$.

\begin{table*}[t]
	\centering
	\caption{Interatomic distances and angles involving the Br$\cdots$Br bonds at $T = 85$~K. $\tau$ is the dihedral angle. The data separate by distance into three groups of four and one group of two, which correspond to pathways composing the leg, rung, diagonal, and interladder interactions. The different interaction parameters appearing in a minimal Heisenberg spin Hamiltonian (Fig.~\ref{fig:jcouplings}) are given in the ``Interaction'' column. 
 Entries in bold text show the most significant discrepancies between the inequivalent ladders of the low-temperature structure (because of their long bond lengths, we do not denote the discrepancies in the diagonal and interladder pathways as ``significant'').
	\label{tab:cpa_contacts}}
	\begin{tabular}{@{}l l l l l l l l l l@{}}
		\toprule
        $\;\;$ \multirow{2}{*}{Interaction} & $\;\;$ \multirow{2}{*}{Atoms} & \multicolumn{4}{c}{$T = 85$~K} & \multicolumn{4}{c}{$T = 203$~K}\\
        & & $d_{\mathrm{Br}-\mathrm{Br}}$ [\AA] & $\theta_{\mathrm{Cu}-\mathrm{Br}\cdots \mathrm{Br}}$ [°]& $\theta_{\mathrm{Br}\cdots \mathrm{Br}-\mathrm{Cu}}$ [°]& $\tau$ [°] &  $d_{\mathrm{Br}-\mathrm{Br}}$ [\AA] & $\theta_{\mathrm{Cu}-\mathrm{Br}\cdots \mathrm{Br}}$ [°]& $\theta_{\mathrm{Br}\cdots \mathrm{Br}-\mathrm{Cu}}$ [°]& $\tau$ [°]\\
        \midrule
        $\;\;$ $J_{\mathrm{leg},1a}$ & $\;\;$Cu$_{1a}$--Br$_{1a}\cdots$Br$_{3a}$--Cu$_{1a}$ & 3.856(3) & 149.6(1) & 149.8(1) & 65.2(3) &  \multirow{2}{*}{3.893(3)} & \multirow{2}{*}{149.8(3)} & \multirow{2}{*}{150.1(1)} & \multirow{2}{*}{55.9(5)} \\
        $\;\;$ $J_{\mathrm{leg},1b}$ & $\;\;$Cu$_{1b}$--Br$_{1b}\cdots$Br$_{3b}$--Cu$_{1b}$ & 3.871(3) & 150.0(1) & 149.6(1)  & 58.8(4) & \\
        $\;\;$ $J_{\mathrm{leg},2a}$ & $\;\;$Cu$_{2a}$--Br$_{5a}\cdots$Br$_{7a}$--Cu$_{2a}$ & 3.841(3) & \bf{148.4(1)} & \bf{151.5(1)} & 66.9(3) &  \multirow{2}{*}{3.881(3)} & \multirow{2}{*}{\bf 148.8(1)} & \multirow{2}{*}{\bf 151.0(2)} & \multirow{2}{*}{64.3(5)}\\
        $\;\;$ $J_{\mathrm{leg},2b}$ & $\;\;$Cu$_{2b}$--Br$_{5b}\cdots$Br$_{7b}$--Cu$_{2b}$ & 3.849(3) & \bf{149.8(1)} & \bf{149.2(1)} & 71.4(3) &  \\
        \\
        $\;\;$ $J_{\mathrm{rung},a}$ & $\;\;$Cu$_{1a}$--Br$_{1a}\cdots$Br$_{8a}$--Cu$_{2a}$ & \bf{4.350(4)} & 100.7(1) & \bf{134.5(1)} & 65.8(2) &  \multirow{2}{*}{\bf 4.396(4)} & \multirow{2}{*}{105.3(1)} & \multirow{2}{*}{\bf 133.8(2)} & \multirow{2}{*}{62.5(3)}\\
        $\;\;$ $J_{\mathrm{rung},b}$ & $\;\;$Cu$_{1b}$--Br$_{1b}\cdots$Br$_{8b}$--Cu$_{2b}$ & \bf{4.526(5)} & 101.3(1) & \bf{131.5(1)} & 66.4(2) & \\
        $\;\;$ $J_{\mathrm{rung},a}$ & $\;\;$Cu$_{1a}$--Br$_{2a}\cdots$Br$_{7a}$--Cu$_{2a}$ & \bf{4.408(3)} & 133.3(1) & 99.9(1) & 66.9(2) &  \multirow{2}{*}{\bf 4.519(4)} & \multirow{2}{*}{132.1(2)} & \multirow{2}{*}{103.2(1)} & \multirow{2}{*}{63.4(2)}\\
        $\;\;$ $J_{\mathrm{rung},b}$ & $\;\;$Cu$_{1b}$--Br$_{2b}\cdots$Br$_{7b}$--Cu$_{2b}$ & \bf{4.534(3)} & 133.2(1) & 99.9(1)& 66.9(2) & \\
        \\
        $\;\;$ $J_{\mathrm{diag},a}$ & $\;\;$Cu$_{1a}$--Br$_{2a}\cdots$Br$_{5a}$--Cu$_{2a}$ & 4.960(3) & 114.1(1) & 145.1(1) & 22.7(2) &  \multirow{2}{*}{4.930(4)} & \multirow{2}{*}{114.6(1)} & \multirow{2}{*}{143.6(1)} & \multirow{2}{*}{27.1(4)}\\
        $\;\;$ $J_{\mathrm{diag},b}$ & $\;\;$Cu$_{1b}$--Br$_{2b}\cdots$Br$_{5b}$--Cu$_{2b}$ & 5.078(3) & 113.3(1) & 146.3(1) & 28.4(2) & \\
        $\;\;$ $J_{\mathrm{diag},a}$ & $\;\;$Cu$_{1a}$--Br$_{3a}\cdots$Br$_{8a}$--Cu$_{2a}$ & 4.953(3) & 145.8(1) & 113.8(1) & 23.2(2) & \multirow{2}{*}{4.919(4)} & \multirow{2}{*}{143.8(1)} & \multirow{2}{*}{114.6(1)} & \multirow{2}{*}{24.6(4)}\\
        $\;\;$ $J_{\mathrm{diag},b}$ & $\;\;$Cu$_{1b}$--Br$_{3b}\cdots$Br$_{8b}$--Cu$_{2b}$ & 5.161(3) & 144.2(1) & 112.7(1) & 25.3(2) & \\
        \\
        $\;\;$ $J_\mathrm{interladder,1}$ & $\;\;$Cu$_{1a}$--Br$_{4a}\cdots$Br$_{6b}$--Cu$_{2b}$ & 5.080(4) & 107.2(1) & 90.9(1) & 173.4(1) & \multirow{2}{*}{5.146(5)} & \multirow{2}{*}{94.3(1)} & \multirow{2}{*}{107.0(1)} & \multirow{2}{*}{170.5(3)}\\
        $\;\;$ $J_\mathrm{interladder,2}$ & $\;\;$Cu$_{1b}$--Br$_{4b}\cdots$Br$_{6a}$--Cu$_{2a}$ & 4.956(4) & 105.7(1) & 92.5(1) & 170.0(1) & \\
	    \bottomrule
    \end{tabular}
\end{table*}

To address the {Cu--Cu} pathways, in Table~\ref{tab:cpa_contacts} we show the {Br$\cdots$Br} distances, {Cu--Br$\cdots$Br} and {Br$\cdots$Br--Cu} angles, and the dihedral angle $\tau$ for each of the inequivalent ladders. We denote sites in the two inequivalent ladders with the subscripts $a$ and $b$. The shortest Br$\cdots$Br distances correspond to the ladder legs, and are shortened by at most 1.7\% on passing from the 203~K structure to the 85~K structure, while the Cu--Br$\cdots$Br angles change by at most 0.7\%. A more pronounced change is found on the ladder rungs (second group of four in Table~\ref{tab:cpa_contacts}), where some bond angles decrease by up to 4\% in both ladder~$a$ and ladder~$b$. On the intraladder, diagonal pathway (third group of four), the halogen-bond length increases in both inequivalent ladders, by 1\% to 4\%. If one inquires about the biggest change between the two inequivalent ladders, this is found in the halogen-bond length on the rungs, which increases by up to 1.5\% on ladder~$a$ while decreasing by up to 1.8\% on ladder~$b$. 

%FIG XX - interactions
\begin{figure}[b]
\centering
\includegraphics[width={\columnwidth}]{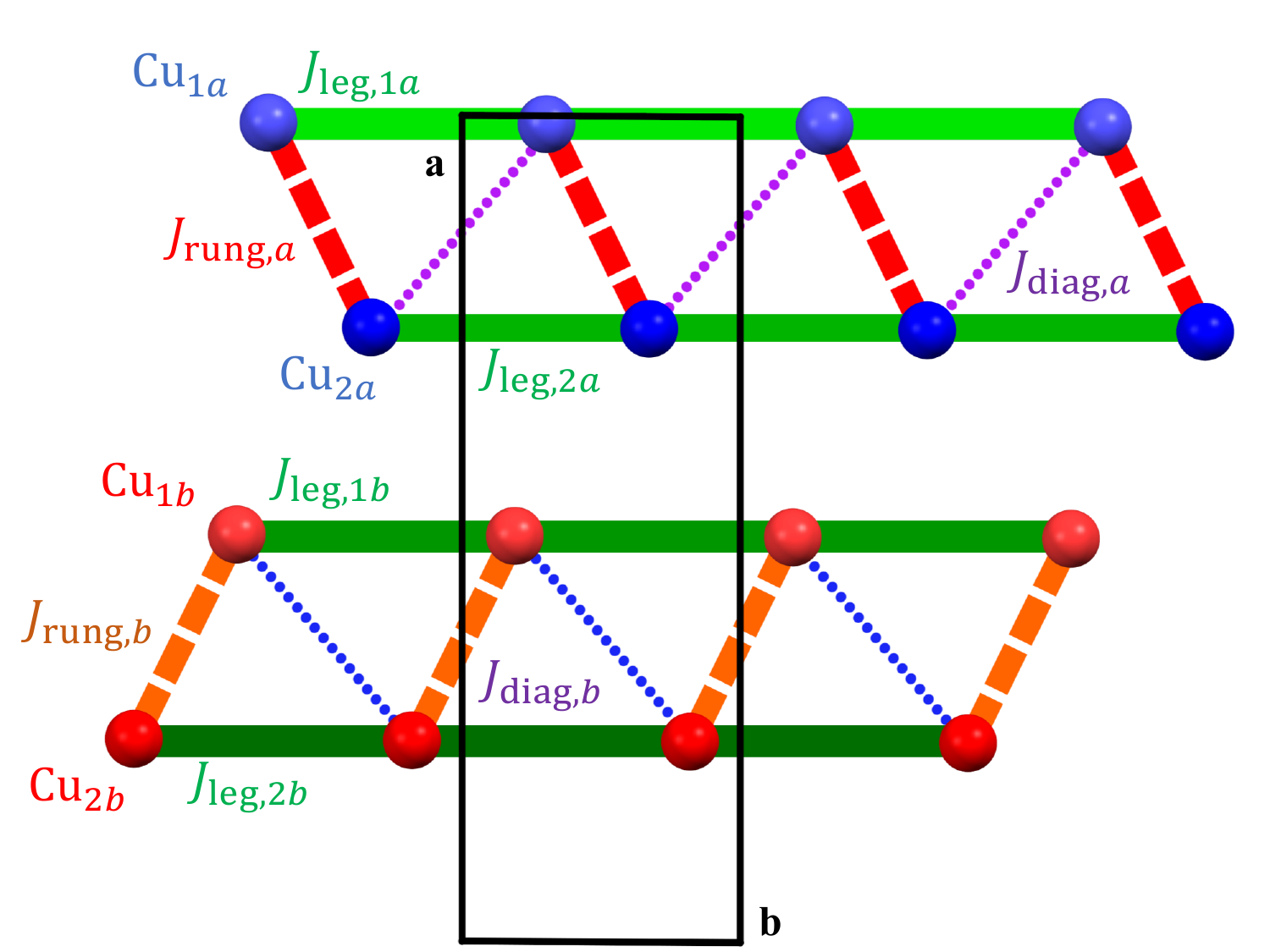}
\caption{\label{fig:jcouplings}
Schematic representation of the interaction parameters expected on the basis of the inequivalent atomic pathways in the two ladders of \CPA\ (Table~\ref{tab:cpa_contacts}). The four inequivalent leg interactions, $J_{\mathrm{leg},1a}$, $J_{\mathrm{leg},2a}$, $J_{\mathrm{leg},1b}$, and $J_{\mathrm{leg},2b}$, are shown as green, solid lines. The two inequivalent rung interactions, $J_{\mathrm{rung},a}$ and $J_{\mathrm{rung},b}$, are shown respectively as dashed red and orange lines. The diagonal interactions, $J_{\mathrm{diag},a}$ and $J_{\mathrm{diag},b}$ are depicted respectively as purple and blue dotted lines. A possible small interladder interaction is not represented.}
\end{figure}

\subsection{Spin Hamiltonian \label{ssec:sh}}

In an ideal two-leg ladder one expects that the spin Hamiltonian contains only two interactions, the Heisenberg superexchange terms $J_\mathrm{leg}$ and $J_\mathrm{rung}$. Here we note that in \CPA\ even the 203~K structure may have two additional complications due to the fact that the two Cu atoms are inequivalent. First, the two leg bonds may not be identical. Second, the rungs possess no center of inversion symmetry and hence may have a finite DM interaction; a rung DM term can cause significant modification of the magnetic properties of a ladder \cite{Penc2007}, for which it is also more effective than a leg DM term. As noted in Sec.~\ref{sec:intro}, DM effects have been documented in near-ideal ladder materials including BPCB and DIMPY \cite{Krasnikova_2020,Ozerov_2015_ESR,Glazkov_2015_ESR}. 

At low temperatures, the number of interaction parameters required to describe Cu-CPA is doubled. Because the Cu--Br$\cdots$Br--Cu pathways differ slightly in length and angle for each of the four inequivalent copper sites, we expect four different values for the leg interactions in \CPA\ (Table \ref{tab:cpa_contacts}, first group of four), as depicted in Fig.~\ref{fig:jcouplings}. Because the geometries of all four pathways remain rather similar at low temperatures, one may anticipate that the corresponding interaction parameters should be comparable in strength. However, the extreme sensitivity of superexchange interactions to bond lengths and angles means that our results certainly do not exclude differences in the range of 10s of percent. 

Turning to the rung pathways (Table \ref{tab:cpa_contacts}, second group of four), it is clear that \CPA\ features two sets of rung interaction parameters. From the bond lengths and angles highlighted in bold text, these should be more different from one another than any of the other parameter groups. Thus \CPA\ could offer a superposition of two strong-leg ladders whose leg-to-rung coupling ratios differ significantly, due primarily to differences in $J_\mathrm{rung}$. 

All further Br$\cdots$Br bonds are considerably longer again (close to 5 \AA), suggesting very small interactions. These fall into two groups, the first corresponding to one diagonal intraladder interaction per inequivalent ladder (Table \ref{tab:cpa_contacts}, third group of four); at lowest order, a diagonal term in a two-leg ladder is an interrung interaction whose effective sign is opposite to $J_\mathrm{leg}$, and in \CPA\ should be a negligible alteration to the effects of the four $J_\mathrm{leg}$ bonds. The second group (Table \ref{tab:cpa_contacts}, final pair) corresponds to interladder bonds, which even if tiny would dictate the onset of three-dimensional magnetic order in applied magnetic fields above the ladder gap, or gaps \cite{Klanjsek2008,Thielemann2009}.

\section{Magnetic properties\label{sec:magprop}}

We have performed a number of measurements of the low-temperature magnetic properties with a view to understanding whether the differences in interaction parameters may be discernible. We remark at the outset that the authors of Ref.~\cite{Willett2004} were able to reproduce their observed magnetic susceptibility using the minimal model of an ideal quantum spin ladder with only two parameters, $J_\mathrm{leg}$ and $J_\mathrm{rung}$. Hence it is possible that the differences between magnetically inequivalent ladders may simply be too small to matter, and we will attempt to gauge this situation in what follows. 

%==== figure =============================%
\begin{figure}[p]
\centering
\includegraphics[width={\columnwidth}]{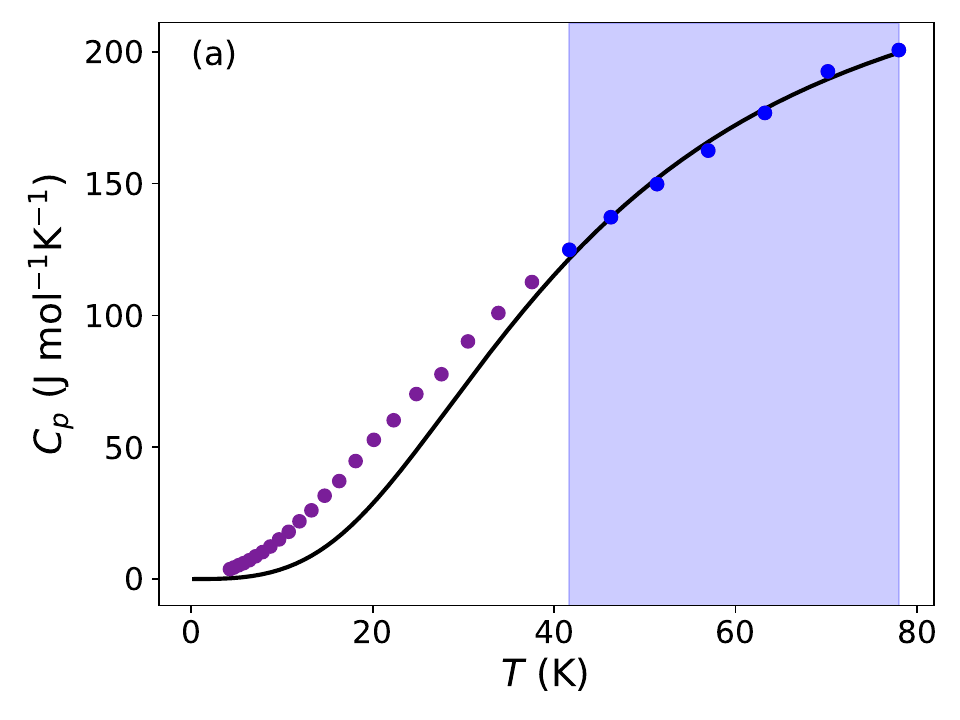}
\includegraphics[width={\columnwidth}]{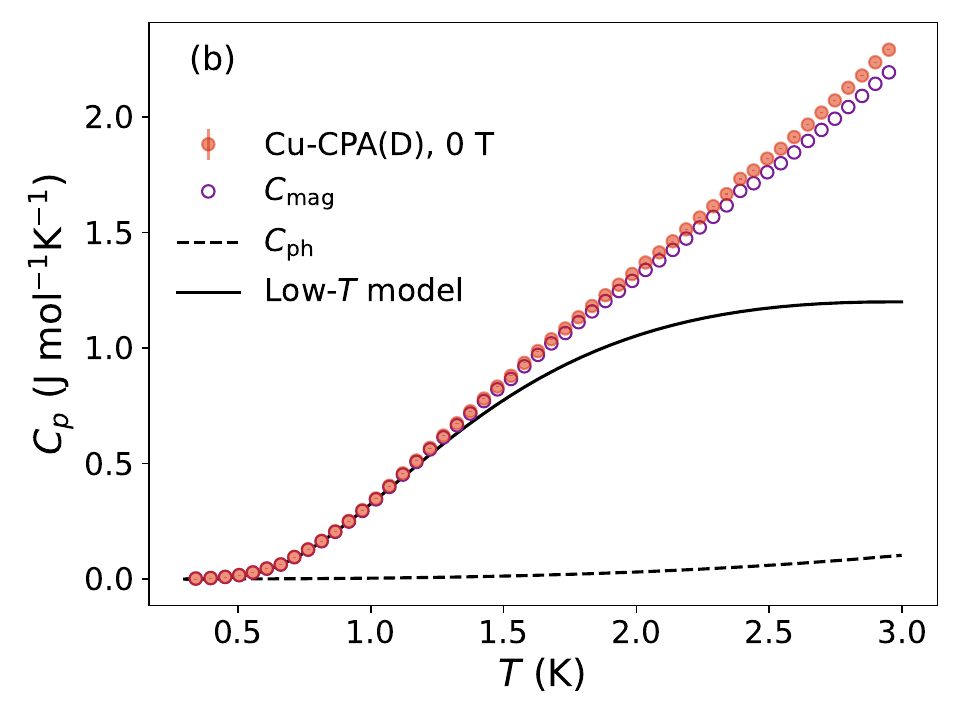}
\includegraphics[width={\columnwidth}]{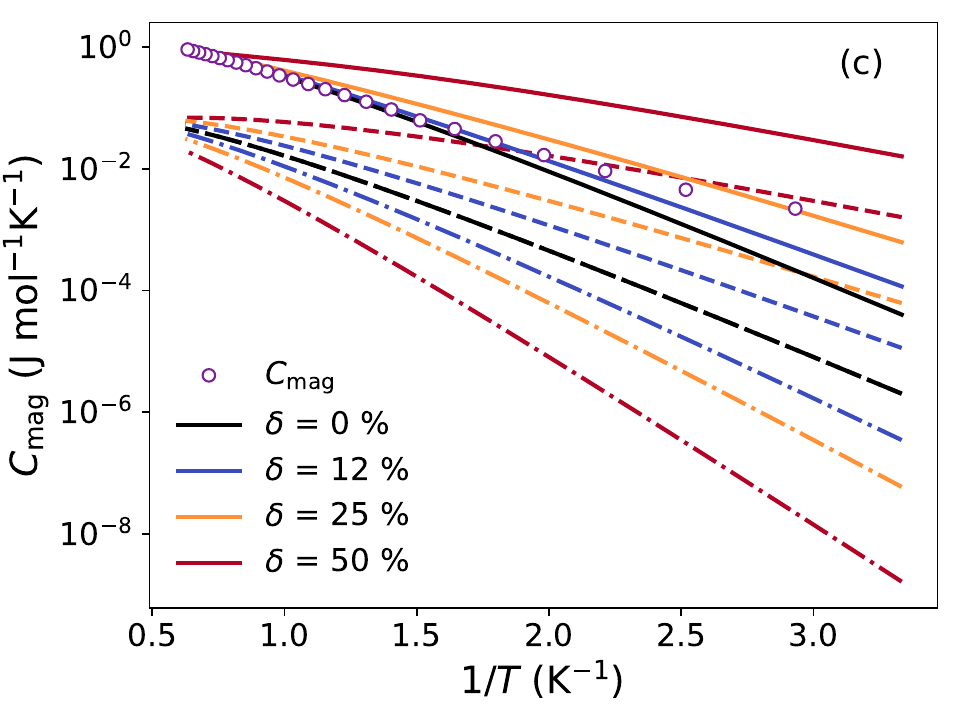}
\caption{\label{fig:CP_fit_low-T}
(a) Zero-field specific-heat of our deuterated sample from Fig.~\ref{fig:heatcapacity}, shown for temperatures up to 80 K. A fit (solid black line) of the data between 40 and 80 K (blue shading) to the standard Debye form, yields a characteristic temperature $T_{\rm D} = 173(4)$ K and prefactor $C_{\rm D} = 84(2)$~JK$^{-1}$mol$^{-1}$. (b) Low-temperature specific heat from panel (a), showing the magnetic contribution ($C_\mathrm{mag}$) obtained by subtraction of the phonon part ($C_\mathrm{ph}$) and a fit of the low-temperature data ($T \le 1.5$ K) to the form of Eq.~\eqref{eq:cgm}. Fitting to a single ladder yields a gap $\Delta = 0.37(2)$~meV and a velocity parameter $c = 0.87(2)$~meV. (c) Test of the low-temperature specific heat for a system composed of ladders with two different gaps. Dashed and dot-dashed lines show respectively the contributions of ladders with smaller and larger gaps $\Delta_\pm = \Delta (1 \pm \delta)$. Solid lines show the sum of these two contributions. The experimental data (open circles) show a systematic deviation from the expected straight-line form in the $T \rightarrow 0$ limit, but are best fitted by a discrepancy parameter $\delta = 13.5$\%.}
\end{figure}

\subsection{Magnetic specific heat\label{ssec:low-T-HC}}

We begin by analyzing our low-temperature specific-heat measurements in order to isolate the magnetic contribution and to test whether it allows the extraction of one, or possibly two distinct, spin gap(s). At temperatures sufficiently far below the structural phase transitions at $T^\star$ and $T^\mathrm{mono}$, a regime we adjudge to be below 80 K, we assume that the specific heat consists of only two contributions, $C_p = C_{\rm mag} + C_{\rm ph}$, from the magnetic sector and from the lattice. To separate the two terms, we note that the high energy scale ($J_\mathrm{rung} + 2J_\mathrm{leg}$) of the assumed single ladder \cite{Willett2004} is of order 30 K, and thus we assume that $C_p$ in the range from 40 to 80 K is almost exclusively phononic. A fit to the standard Debye form, shown in Fig.~\ref{fig:CP_fit_low-T}(a), substantiates these assumptions and returns an estimate of the Debye temperature as $T_{\rm D} = 173(4)$ K, with prefactor $C_{\rm D} = 84(2)$~JK$^{-1}$mol$^{-1}$. In the truly low-$T$ regime of interest, shown in Fig.~\ref{fig:CP_fit_low-T}(b), only the acoustic phonons contribute and the lattice contribution has the well known form $C_\mathrm{ph} = bT^3$, but $C_{\rm ph}$ is so much smaller than $C_{\rm mag}$ that an accurate estimate of the coefficient $b$ is clearly not necessary. 

The wide band of spin excitations in the strong-leg ladder ensures that $C_{\rm mag}$ has significant contributions over a broad range of energies, such that quantitatively accurate modeling is a complex process. We therefore restrict our considerations to temperatures well below the spin gap, where we apply the expression deduced \cite{Troyer1994} for a 1D gas of particles with acoustic dispersion $E = \sqrt{\Delta^2 + c^2 k^2}$, 
\begin{equation}
\label{eq:cgm}
\!\!\!\! C_\mathrm{mag} = \frac{3R}{2\sqrt{2\pi}} \! \left( \! \frac{\Delta}{k_\mathrm{B}T} \!\! \right)^{\frac{3}{2}} \!\! \frac{\Delta}{c} \! \left[ \! 1 \! + \! \frac{k_\mathrm{B}T}{\Delta} \! + \! \frac{3}{4} \! \left( \! \frac{k_\mathrm{B}T}{\Delta} \! \right)^2 \right] e^{-\frac{\Delta}{k_\mathrm{B}T}}. 
\end{equation}
Quantitatively, fitting to temperatures $T < \Delta/4$ justifies using only the first term of Eq.~\eqref{eq:cgm} \cite{Hong2010}, while retaining all three terms extends the validity range to approximately $T < \Delta/2$. Fitting the data below 1.5 K to a single ladder, shown in Fig.~\ref{fig:CP_fit_low-T}(b), returns the parameters $\Delta = 0.37(2)$~meV and $c = 0.87(2)$~meV; we remark that such a direct measurement of the spin gap was not previously available for Cu-CPA.

Armed with the knowledge that Cu-CPA is composed of two potentially quite different spin ladders with equal volume fractions, the model of Eq.~\eqref{eq:cgm} offers the possibility of testing how different discrepancies between ladder parameters would appear in a thermodynamic property such as the low-$T$ specific-heat. As a preliminary step in this direction, in Fig.~\ref{fig:CP_fit_low-T}(c) we show the specific-heat contributions in the low-$T$ limit from two ladders with the same $c$ parameter, but whose gaps take the values $\Delta_\pm = \Delta (1 \pm \delta)$. As the single discrepancy parameter, $\delta$, is increased, the ladder with the smaller gap plays an increasingly dominant role at truly low temperatures due to the exponential term, but the recovery arising from the power-law terms suggests already that Eq.~\eqref{eq:cgm} is reaching its limits at the left-hand side of Fig.~\ref{fig:CP_fit_low-T}(c). Adding the two contributions [solid lines in Fig.~\ref{fig:CP_fit_low-T}(c)] leads to the conclusion that significant $\delta$ values would indeed be discernible in measurements of $C_{\rm mag}$. 

Overplotting our own data reveals a small departure from the expected linear form at the lowest temperatures that is strongly magnified by the semi-log and inverse-temperature axes, and is presumably a consequence of impurities that were not included in the two-component model. One may deduce that the gap estimate is provided by fitting the data in the range $0.5 < T < 1.5$~K. The best fit under the circumstances is provided by $\delta = 13.5\%$, very close to the 12\% curve shown, which would imply the two gaps $\Delta_1 \equiv \Delta_- = 0.32$ meV and $\Delta_2 \equiv \Delta_+ = 0.42$ meV. We stress again that we provide this analysis largely to illustrate the effect of two inequivalent ladders, and not as a quantitative claim concerning either the discrepancy or the gaps. Such a discrepancy is nevertheless eminently reasonable on the basis of the considerations in Sec.~\ref{sec:magint}. By contrast, the gaps we deduce are very much larger than those deduced from the susceptibility, as we explain next, and we point again to the need for spectroscopy experiments that will provide a definitive answer to the Hamiltonian parameters for Cu-CPA. 

\subsection{Other magnetic properties}

We have also measured the magnetic susceptibility and obtained results fully consistent with those of Ref.~\cite{Willett2004}. As already noted, no direct signature of the two-ladder nature of \CPA\ can be found in these data, and we show this explicitly in App.~\ref{appb}. Quite generally, the magnetic susceptibility, and indeed most other bulk quantities, provide very general information from which it is possible to infer only a small number of independent energy scales. This is particularly true in \CPA, where the rather weak interactions deduced from the susceptibility, $J_\mathrm{leg} = 1.0$~meV and $J_\mathrm{rung} = 0.47$~meV, mandate dilution temperatures to extract the spin gap directly. As noted above, these interaction parameters imply a spin gap, 0.20 meV \cite{Willett2004}, that is little over half of the single gap we extract directly from the specific heat, which implies that a reassessment of either the spin Hamiltonian or the value of the fitting technique may be in order. 

Nevertheless, explicit formulas exist for the susceptibility of a two-leg ladder, and in App.~\ref{appc} we use these to illustrate the ways in which inequivalent leg and rung interactions would become observable.
From the resulting observation that the respective interactions must differ by many tens of percent in order to become detectable, we conclude again that detailed spectroscopic studies, preferably in combination with \textit{ab initio} calculations, are required for the systematic determination of the multiple interaction parameters in Fig.~\ref{fig:jcouplings}.

We turn next to measurements of the magnetization, $M(H)|_T$. Although the results of Ref.~\cite{Willett2004} were obtained at temperatures below the estimated spin gap, and show the expected trend of a monotonic increase until saturation, they show neither a sharp onset at low fields nor the approach to saturation of a ladder model. With the insight that the system possesses two inequivalent ladders, the slow onset of the magnetization may be explained by the presence of two spin gaps. Similarly, departure of the near-saturation behavior from that of one ideal ladder could be a result of two distinct saturation fields. 
Specifically, at low $T$ an ideal ladder shows a sharp increase in $M(H)$ up to the saturation plateau, and a mismatch in saturation fields between the two different ladders would create two such steps, broadening the expected feature.
Finally, it has been proposed that subtle tendencies toward the formation of plateaus in the magnetization could appear due to the field-induced reorientation of the CuBr$_{4}^{2-}$ anions \cite{Woodward2005}. While this physics may occur, a detailed understanding of the spin excitation spectrum is required to exclude simpler scenarios. In the light of our structural findings, a sub-100~mK magnetization measurement should be combined with neutron spectroscopy to elucidate the magnetic behavior of \CPA.

An important implication of the rather strong isotope effect we observe on the phase-transition temperatures is the softness of the structure. As a result, \CPA\ is an excellent candidate for studying pressure-induced quantum phenomena \cite{Sachdev2008}. In contrast to previous studies of pressure effects in dimerized quantum magnets, which concentrated on strong-dimer materials with well defined triplon or magnon excitations \cite{Ruegg2004,Thede2014,Bettler2017}, \CPA\ is thought to realize a model displaying the fingerprints of weakly confined spinons, and hence offers different possibilities for quantum phase transitions and the evolution of spin excitations. Finally, it remains to be determined whether the excitations of the two inequivalent ladders may have an appreciable interaction, which could manifest itself in unconventional behavior either under ambient conditions or under applied external fields or pressures. 

\section{Conclusions \label{sec:conclusion}}

In summary, our studies reveal two previously unknown structural phase transitions in the metal-organic quantum magnet \chemCPA~(Cu-CPA). By characterizing the low-temperature structure we establish \CPA\ as an experimental realization of a two-ladder model and hence as a promising material in which to search for additional magnetic excitations in the strong-leg regime. 
We use our own thermodynamic measurements for a direct measurement of the spin gap, or gaps, and to illustrate the consequences of two-ladder character in the magnetic properties. 

\begin{acknowledgments} 
We thank D.~Le and R.~Valenti for helpful discussions. We are grateful to T.~Greber for enabling us to measure d.c.~susceptibilities on an MPMS3 instrument that was acquired as part of the SNSF R'Equip Grant No.~ASKUZI 206021\_150784. This project was supported by the Horizon 2020 research and innovation program of the European Union under the Marie Skłodowska-Curie Grant (Agreement No.~884104, PSI-FELLOW-III-3i). 
We acknowledge in addition the support of the UZH Candoc/Postdoc Grant 2023 Verfügung~Nr.~FK-23-107 and 
of the Chalmers X-Ray and Neutron Initiatives (CHANS) and the Swedish Research Council (VR) through a 
Starting Grant (No. Dnr.~2017-05078), through Grant Nos. Dnr.~2022-06217, Dnr.~2021-06157, and Dnr.~2022-03936, 
through the Foundation Blanceflor 2023 fellow scholarship, and through the Area of Advance-Material Sciences of Chalmers University of Technology. We are grateful to the Paul Scherrer Institute for the allocation of neutron beam time on CAMEA at SINQ (under Project No.~20212923). 
\end{acknowledgments} 

\begin{appendix}

\section{Structural properties\label{sec:structural_properties}}

We summarize the properties of \chemCPA\ and \chemCPAD\ in Table~\ref{tab:chemical_properties}. We draw attention to the mass difference, which is important for an accurate comparison of respective specific-heat measurements. 

%==== figure =============================%
\begin{figure}[t]
\centering
\includegraphics[width={\columnwidth}]{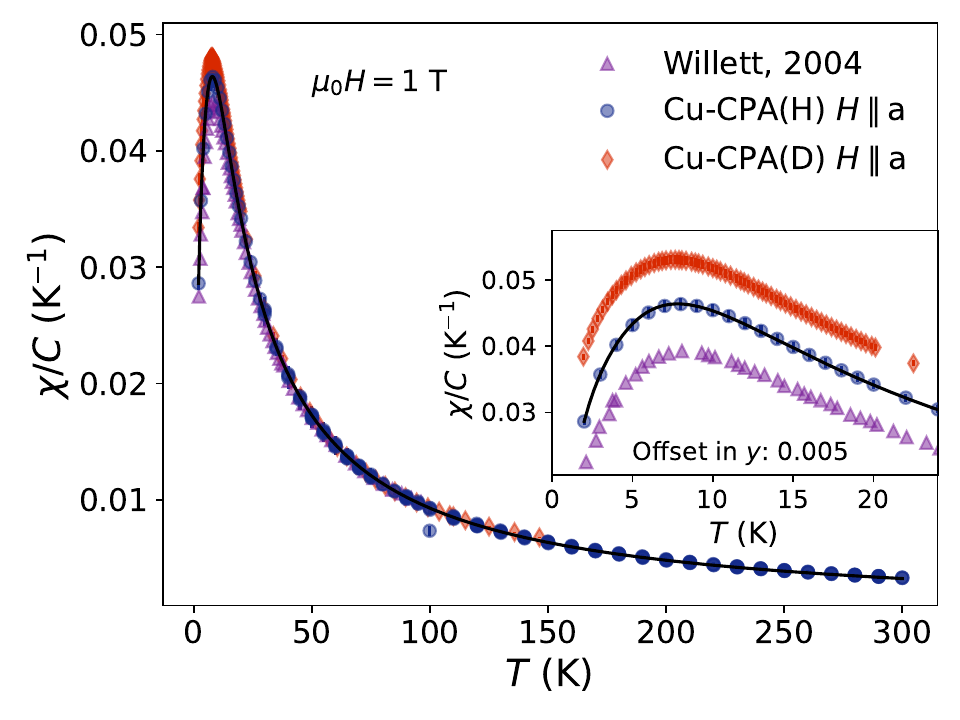}\\
\caption{\label{fig:susceptibility_normalized_willett}
Normalized magnetic susceptibility of our hydrogenated (blue circles) and deuterated (red diamonds) samples, shown together with the measurements of Ref.~\cite{Willett2004} (purple triangles). The solid black line represents the best fit to Eq.~\eqref{eq:chi_strong_leg}, which returns the parameters shown in Table \ref{tab:chi_fit_results_P-C}.}
\end{figure}

\begin{table}[b]
\caption{\label{tab:chemical_properties}
Structural properties of the hydrogenated and deuterated compounds, obtained respectively from powder X-ray diffraction at 95 and single-crystal X-ray diffraction at 85~K.}
\begin{tabular}{@{}l l l@{}}
\toprule
$\;\;$ chemical formula $\;\;\;\;$ & C$_{10}$H$_{24}$Br$_{4}$CuN$_{2}$ $\;\;\;\;$ & C$_{10}$D$_{24}$Br$_{4}$CuN$_{2}$ $\;\;$ \\
\hline
$\;\;$ f.w. [g/mol] & 555.49 & 579.29 \\
$\;\;$ T [K] & 95(2) & 85(2)\\
$\;\;$ cryst. system & monoclinic & monoclinic\\
$\;\;$ space group & $P112_1$ & $P112_1$\\
$\;\;$ $a$ [\AA] & 23.9649(1) & 23.9549(16)\\
$\;\;$ $b$ [\AA] & 8.0765(1) & 8.0759(6) \\
$\;\;$ $c$ [\AA] & 18.2819(1) & 18.2733(13) \\
$\;\;$ $\alpha$ [°] & 90 & 90 \\
$\;\;$ $\beta$ [°] & 90 & 90\\
$\;\;$ $\gamma$ [°] & 90.35(1) & 90.30(1)\\
$\;\;$ V [\AA${}^{3}$] & 3538.44 & 3535.06\\
\bottomrule
\end{tabular}
\end{table}

\section{Magnetic susceptibility}
\label{appb}

Because the bulk magnetic properties of multiple gapped and low-dimensional quantum magnets are qualitatively similar \cite{Landee2014}, it is desirable to have an accurate quantitative model for any quantity. We have measured the magnetic susceptibility for comparison with the highly accurate polynomial expansion obtained from quantum Monte Carlo (QMC) simulations of the model describing spin-$1/2$ antiferromagnetic Heisenberg ladders \cite{Johnston2000} in the strong-leg ($\alpha \gg 1$) regime,
\begin{equation}
\label{eq:chi_strong_leg}
\chi = \frac{C \! \exp \! \left( \! \frac{-\Delta(\alpha)}{T} \! \right)}{T} \! \frac{1 \! + \! \sum_{i=1}^6 \! \left( \! \frac{J_\mathrm{leg}}{T} \! \right)^i \! \sum_{j=1}^{3} \! N_{i,j} \alpha^{-j}}{1 \! +\! \sum_{i=1}^6 \! \left( \! \frac{J_\mathrm{leg}}{T} \! \right)^i \! \sum_{j=1}^{3} D_{i,j} \alpha^{-j}} \! + \! \frac{PC}{T}.
\end{equation}
Here $\Delta(\alpha)$ is the spin gap, $P$ the concentration of paramagnetic impurities, and $C$ the standard Curie constant, while $N_{i,j}$ and $D_{i,j}$ are coefficients determined by QMC and tabulated in Ref.~\cite{Johnston2000}. For a direct comparison with the model, we also subtracted the diamagnetic contribution. By contrast, the impurity contribution was not directly evident at any of the temperatures accessed in our measurement, meaning that it is reassuringly small, and thus was included in the primary fit. 

%==== figure =============================%
\begin{figure}[t]
\centering
\includegraphics[width={\columnwidth}]{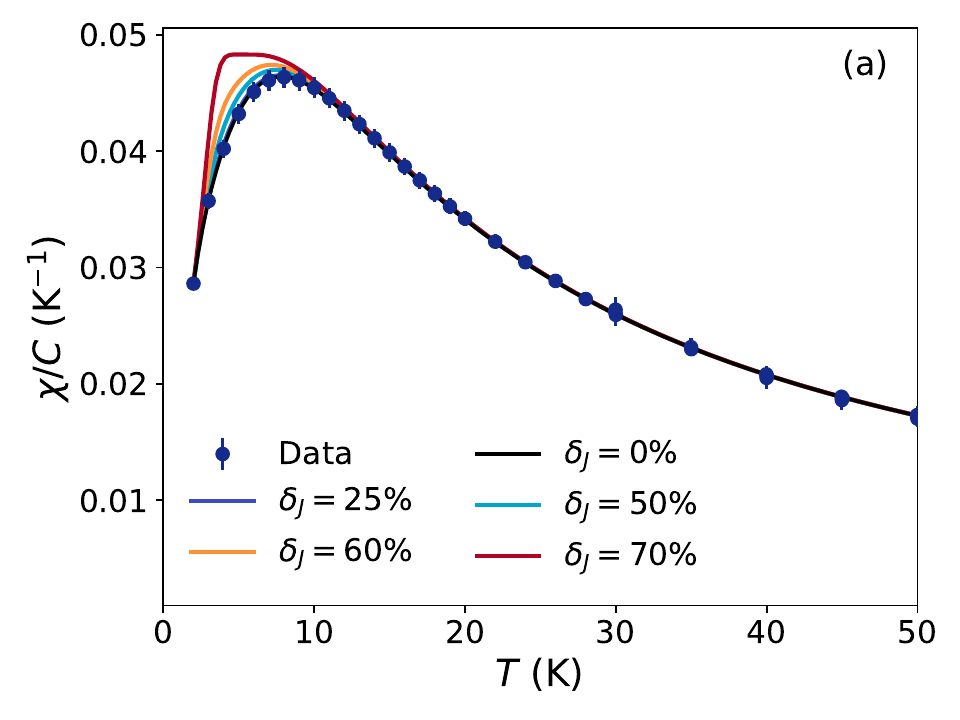}\\
\includegraphics[width={\columnwidth}]{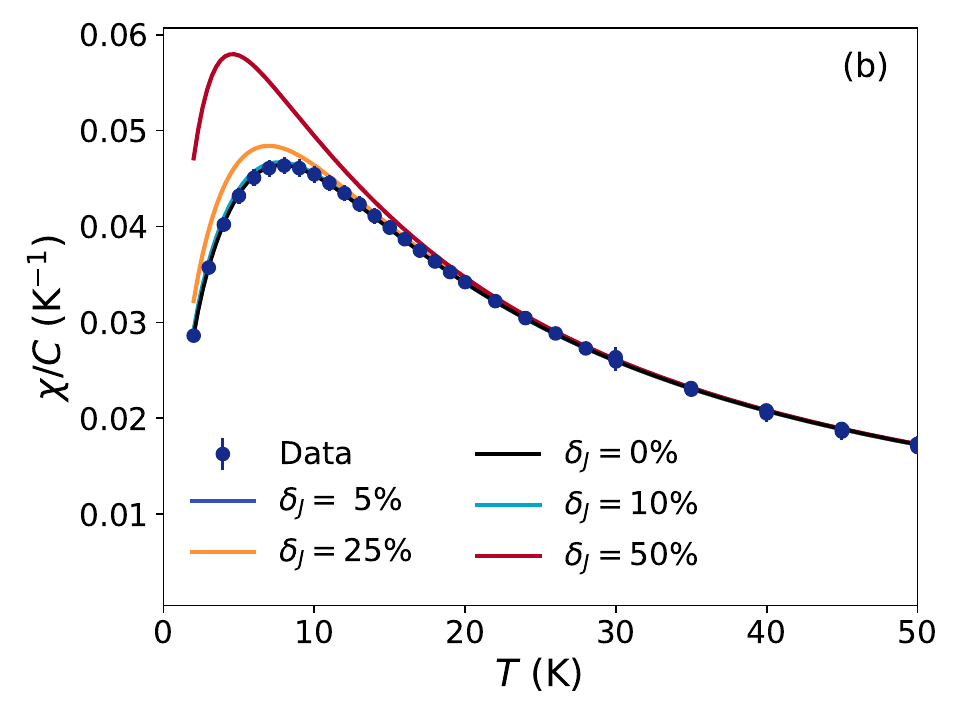}
\caption{\label{fig:2ladders_Jrung}
Normalized magnetic susceptibilities calculated for a system of two inequivalent ladders. (a) Discrepancy only in the rung interactions, $J_\mathrm{rung}^{(1)} = J_\mathrm{rung} (1 - \delta_J)$ and $J_\mathrm{rung}^{(2)} = J_\mathrm{rung} (1 + \delta_J)$. (b) Discrepancy only in the leg interactions, $J_\mathrm{leg}^{(1)} = J_\mathrm{leg} (1 - \delta_J)$ and $J_\mathrm{leg}^{(2)} = J_\mathrm{leg} (1 + \delta_J)$. Symbols mark again the susceptibility measured for our hydrogenated Cu-CPA sample and black lines the best fit to Eq.~\eqref{eq:chi_strong_leg} obtained using a single ladder, equivalent to $\delta_J = 0$.}
\end{figure}

\begin{table}[b]
\centering
\caption{\label{tab:chi_fit_results_P-C}
Parameters fitting the magnetic susceptibility as obtained from Eq.~\eqref{eq:chi_strong_leg}.}
%\begin{tabular}{@{}l l l@{}}
\begin{tabular}{@{}l c c}
\toprule
& \chemCPA $\;\;\;\;$ & \chemCPAD $\;\;$\\
\hline
$\;\;$ $C$ [emu~K] $\;\;$ & 0.362(1)$\;\;$  & 0.453(2)$\;\;$ \\ 
$\;\;$ $J_\mathrm{leg}$ [meV] $\;\;$ & $0.999(2)$$\;\;$ & $1.002(3)$ $\;$ \\
$\;\;$ $J_\mathrm{rung}$ [meV] $\;\;$ & $0.458(15)$ & $0.461(10)$ \\  
$\;\;$ $\Delta$ [meV] $\;\;$ & $0.194(7)$ $\;$ & $0.195(5)$ $\;$\\
$\;\;$ $\alpha$ & $2.18(7)$ $\;\;\;$&  $2.17(5)$ $\;\;\;$\\
\bottomrule
\end{tabular}
\end{table}

Our susceptibility data are shown in Fig.~\ref{fig:susceptibility_normalized_willett}. Following Ref.~\cite{Willett2004}, if the 203~K structure is assumed and the susceptibility is fitted to that of a single ladder, one obtains the parameters $J_{\mathrm{leg}}$ and $J_{\mathrm{rung}}$ given in Table~\ref{tab:chi_fit_results_P-C}, and hence the derivative quantities $\Delta$ and $\alpha$, the latter of order 2.2 and hence comfortably in the strong-leg regime. The fit to the data is shown as the solid gray line in Fig.~\ref{fig:susceptibility_normalized_willett}. We remark that the fit of Eq.~\eqref{eq:chi_strong_leg} is very sensitive to the low-temperature data, and that the lowest measured temperature is barely below the spin gap of 0.20 meV that the values of $J_{\mathrm{leg}}$ and $J_{\mathrm{rung}}$ imply. To ensure the validity of the fit, it is helpful to have an independent estimate of the gap, and here we note again the significant mismatch between the value obtained indirectly from $\chi(T)$ and the value of 0.37 meV obtained directly from the specific heat in Sec.~\ref{sec:magprop}A.

\section{Magnetic susceptibility with inequivalent interactions}
\label{appc}

Our discovery of a more complex low-temperature structure paints a picture different from that of the preceding appendix. Like the magnetic specific heat in Subsec.~\ref{ssec:low-T-HC}, the magnetic susceptibility modeled using Eq.~\eqref{eq:chi_strong_leg} offers the possibility of computing the effects of two inequivalent spin ladders. To this end, we define the inequivalent rung and leg interaction parameters for ladders 1 and 2 in terms of a single difference parameter, $\delta_J$, as $J_\mathrm{rung}^{(1)} = J_\mathrm{rung} (1 - \delta_J)$, $J_\mathrm{rung}^{(2)} = J_\mathrm{rung}(1 + \delta_J)$, $J_\mathrm{leg}^{(1)} = J_\mathrm{leg} (1 - \delta_J)$, and $J_\mathrm{leg}^{(2)} = J_\mathrm{leg} (1 + \delta_J)$, where $J_\mathrm{rung}$ and $J_\mathrm{leg}$ are the single-ladder parameters specified in Table~\ref{tab:chi_fit_results_P-C}.

The normalized magnetic susceptibilities calculated for different values of $\delta_J$ are shown in Fig.~\ref{fig:2ladders_Jrung}. We separate the effects of inequivalence into a rung discrepancy only [Fig.~\ref{fig:2ladders_Jrung}(a)] and a leg discrepancy only [Fig.~\ref{fig:2ladders_Jrung}(b)]. We observe that the susceptibility, measured down to 2~K, does not show any deviations as a consequence of setting two different rung interactions until $\delta_J$ is as large as 25\% (i.e.~$J_\mathrm{rung}^{(1)}/J_\mathrm{rung}^{(2)} < 0.6$), although marked features begin to arise at truly large $\delta_J$. By contrast, the susceptibility is somewhat more sensitive to variations of $J_\mathrm{leg}$, with clear differences at $\delta_J = 0.25$ and strong ones when $\delta_J = 0.5$. Presumably for initial parameters with $\alpha > 2$, this is a consequence of the larger $J_{\rm leg}$ interaction causing a significant reduction of the gap in one of the two inequivalent ladders.

\end{appendix}

%\bibliography{CPAbib}

\end{document}